\documentclass[cancers,article,accept,pdftex,moreauthors]{Definitions/mdpi} 

\usepackage{algorithm}
\usepackage{algorithmic}
\usepackage{booktabs}

\firstpage{1} 
\pubvolume{1}
\issuenum{1}
\articlenumber{0}
\pubyear{2023}
\copyrightyear{2023}
\externaleditor{Academic Editor: Firstname Lastname}
\datereceived{ 9 June 2023 } 
\daterevised{ 16 August 2023} 
\dateaccepted{ 17 August 2023} 
\datepublished{ 22 August 2023} 
\hreflink{https://doi.org/10.3390/cancers15174210} 

\Title{Interpreting the Mechanism of Synergism for Drug Combinations Using Attention-Based Hierarchical Graph Pooling } 

\TitleCitation{Interpreting the Mechanism of Synergism for Drug Combinations Using Attention-Based Hierarchical Graph Pooling }

\Author{Zehao Dong 
 $^{1}$, Heming Zhang $^{2}$, Yixin Chen $^{1}$, Philip R.O. Payne $^{2}$ and Fuhai Li $^{2,3,}$* }

\AuthorNames{Zehao Dong, Heming Zhang, Yixin Chen, Philip R.O. Payne and Fuhai Li}

\AuthorCitation{Dong, Z.; Zhang, H.;\linebreak  Chen, Y.; Payne, P.; Li, F.}

\address{%
$^{1}$ \quad Department of Computer Science \& Engineering, Washington University in St. Louis,\linebreak  St. Louis, MO 63130, 
 USA; zehao.dong@wustl.edu 
 ; chen@cse.wustl.edu \\
$^{2}$ \quad Institute for Informatics, Data Science, and Biostatistics, Washington University 
 School of Medicine, Washington University in St. Louis, St. Louis, MO 63110, USA; hemingzhang@wustl.edu; prpayne@wustl.edu  \\ 
$^{3}$ \quad Department of Pediatrics, Washington University School of Medicine, Washington University in St. Louis,\linebreak  St. Louis, MO 63110, USA}

\corres{Correspondence: fuhai.li@wustl.edu}

\simplesumm{ This paper introduces a novel graph neural network (a hierarchical graph pooling model), SANEpool, to effectively detect core sub-networks of significant genes for predicting the synergy score of drug/drug combinations in cancer. SANEpool successfully addresses the limitations of the un-transparency in the prediction process of previous computational AI models for drug synergy prediction, while providing the superior predictive performance than popular baselines on numerous drug-synergy prediction datasets. The success of SANEpool indicates that significant gene-gene interactions and gene-drug interactions play a crucial role in designing powerful deep learning models to provide accurate prediction and to reveal the mechanism of the synergy (MoS). } 

\abstract{
Synergistic drug combinations provide huge potentials to enhance therapeutic efficacy and to reduce adverse reactions. However, effective and synergistic drug combination prediction remains an open question because of the unknown causal disease signaling pathways. Though various deep learning (AI) models have been proposed to quantitatively predict the synergism of drug combinations, the major limitation of existing deep learning methods is that they are inherently not interpretable, which makes the conclusions of AI models untransparent to human experts, henceforth limiting the robustness of the model conclusion and the implementation ability of these models in real-world human--AI healthcare. In this paper, we develop an interpretable graph neural network (GNN) that reveals the underlying essential therapeutic targets and the mechanism of the synergy (MoS) by mining the sub-molecular network of great importance. The key point of the interpretable GNN prediction model is a novel graph pooling layer, a self-attention-based node and edge pool (henceforth SANEpool), that can compute the attention score (importance) of genes and connections based on the genomic features and topology. As such, the proposed GNN model provides a systematic way to predict and interpret the drug combination synergism based on the detected crucial sub-molecular network. Experiments on various well-adopted drug-synergy-prediction datasets demonstrate that (1) the SANEpool model has superior predictive ability to generate accurate synergy score prediction, and (2) the sub-molecular networks detected by the SANEpool are self-explainable and salient for identifying synergistic drug combinations.}


\begin{document}



\section{Introduction\label{sec1}}

Combinatorial drug therapy has been of crucial importance in modern clinical disease treatment and drug discovery \cite{hopkins2008network}. Synergistic drug combination can produce more beneficial combinatorial effects than each constituent, and the synergic behavior always allows the lower doses of the drugs in the combination relative to their individual potencies, thus reducing the induction of drug resistance \cite{podolsky2011combination} and overcoming the side effects \cite{chandrasekaran2016chemogenomics,radic2017combinatorial,o2016unbiased} associated with the high doses of single drug usage. Hence, drug combination therapy provides a greatly promising avenue towards the treatment of the most dreadful multi-factorial diseases, such as cancer \cite{devita1975combination,crino1997cisplatin,carew2008histone,shuhendler2010novel}, diabetes, and bacterial infections. In contrast to the synergism, the therapeutic efficacy of some drug combinations can be simply additive or even sub-additive. As such, there has been growing interest in investigating the synergy mechanism of drug combinations to distinguish the synergistic combinations from non-synergistic ones.

Frequently, the synergy of drug combinations is tested in pre-clinical model environments, such as high-throughput screening (HTS) instruments \cite{mott2015high, griner2014high}, where thousands of combinatorial experiments are simultaneously implemented under actionable hypotheses and conditions to profile the synergism. However, the testing space can be extremely massive due to the large amount of drug combinations, cell lines, dose choices, and patient samples, hence it can be impractical to traverse the whole testing space \cite{holbeck2017national}. Furthermore, the transition from some pre-clinical environments to the clinical practice sometimes can also cause failure \cite{yang2018review}. As such, various computational (AI) models \cite{zhang2021synergistic,janizek2018explainable,chen2016xgboost,preuer2018deepsynergy} are developed to assist the synergy analysis of drug combinations. 

Most computational AI models take massive omics data and chemical structure data as input and then adopt deep learning algorithms to predict the synergy score to determine the presence of the synergism. Several machine learning models, such as TreeCombo~\cite{chen2016xgboost} and random forest \cite{sidorov2019predicting}, build ensemble trees to predict the synergy scores and have achieved impressive results. After that, numerous deep learning models were proposed in the domain to unleash the predictive power of neural networks. A large body of work, including DeepSynergy \cite{preuer2018deepsynergy}, MatchMaker \cite{kuru2021matchmaker}, and CCSynergy \cite{hosseini2023ccsynergy}, shows that wisely combining the drug profiles and gene expression profiles in specific cell lines as input features enables the vanilla multiple layer perceptron (MLP) to accurately predict the synergy scores of drug combinations, while TranSynergy \cite{liu2021transynergy} applies attention-based transformer architecture to boost the prediction performance. On the other hand, SDCNet \cite{zhang2022predicting} and DeepDDS \cite{wang2022deepdds} demonstrate that modeling the connections between drugs and genes can benefit synergy prediction and propose to encode the networks/graphs that consist of genes and drugs through graph neural networks (GNNs).

In addition to the predictive ability, the interpretability of deep learning models is desirable in real-world scenarios like the pharmacy industry and healthcare, as it allows us to incorporate human expertise in decision making to provide more robust conclusions. Currently, limited existing works provide interpretable predictions of drug synergy. TranSynergy~\cite{liu2021transynergy} applies the post hoc interpretation framework \cite{carvalho2019machine} that computes the Shapley value of each gene through GradientExplainer and uses DeepExplainer to characterize its contribution to the final synergy prediction. Though post hoc interpretation mechanisms that produce interpretations after the model creation work well in some cases, the ante hoc interpretable model \cite{burkart2021survey} is still missing in the domain of drug synergism analysis to inject interpretability from the beginning of the model design. Consequently, the objective of this paper is to develop deep learning models to generate accurate and interpretable synergic predictions, and we resort to the graph pooling methodology in graph neural networks (GNNs).

In recent years, GNNs have been the dominant architecture for analyzing graph-structured data, such as social networks \cite{monti2017geometric,ying2018graph}, protein networks \cite{fout2017protein,zitnik2018modeling}, circuit networks~\cite{dong2022pace, dong2022cktgnn}, etc. Most GNNs follow the neighborhood aggregation scheme that updates each node feature by propagating its neighboring node features to its current feature and have achieved impressive results on various graph learning tasks, ranging from node classification \cite{hamilton2017inductive} and link prediction \cite{schutt2017schnet,zhang2018link} to graph classification \cite{dai2016discriminative}. In order to generate a subset of nodes or cluster of nodes for the prediction tasks, several graph-pooling models are proposed. DGCNN \cite{zhang2018end} proposes to sort nodes for pooling according to their structural roles within the graph. However, since it stacks multiple graph convolution layers to propagate information and then globally implement the graph down-sampling via a pooling module, the generated graph representation is inherently flat. In order to extract hierarchical graph representations, DiffPool \cite{ying2018hierarchical} uses different GNNs to separately implement neighborhood aggregation and graph pooling, and it provides a framework to hierarchically pool nodes across a broad set of graphs.


Following this inspiration, we introduce a novel hierarchical graph pooling model, SANEpool (self-attention-based node and edge pool), for the interpretable drug synergism prediction task, which aims to reveal the drug combination synergism by systematically extracting the target gene sub-network that intrigues the synergic behavior. Various medical chemistry research has shown that cancer is driven by genetic and epigenetic alterations, many of which can be mapped into signaling pathways that control the survival and migration/invasion of cancer cells.  As one previous signaling pathway analysis suggests~\cite{sanchez2018oncogenic}, 89$\%$ of tumor samples had at least one driver alteration in one of ten cancer-related signaling pathways that is responsible for tumor development, while 57$\%$ and 30$\%$ had one and multiple potentially druggable targets, respectively. Another example \cite{pan2017synthetic} is that the drug combination of venetoclax and idasanutlin can generate antileukemic efficacy in the treatment of acute myeloid leukemia by inhibiting antiapoptotic Bcl-2 family proteins and activating the p53 pathway at same time. Thus, inhibited signaling targets analysis shows great potential of facilitating drug combination synergism discovery. Following this intuition, each SANEpool layer implements the standard graph convolution layer (GCN) to generate attention features that encode the gene (and drug) information as well as the topology information of the molecular network. Based on these attention features, the probability that a gene or an interaction (gene--gene interaction, gene--drug interaction) will cause the synergy performance is calculated, and then genes and interactions that are unlikely to influence the synergism of drug combination will be filtered out. The proposed model is composed of multiple SANEpool layers and will output the target gene sub-network for interpretable and robust synergy prediction. 

We evaluate our SANEpool model on three popular drug-synergy-prediction datasets, which are constructed upon NCI ALMANAC \cite{sidorov2019predicting}, GDSC (Genomics-of-Drug-Sensitivity-in-Cancer) \cite{jaaks2022effective}, and O'Neil \cite{o2016unbiased} experimental settings. Experimental results demonstrate that the SANEpool model achieves the current state-of-the-art performance for all datasets. Furthermore, through visualizations of the detected target gene sub-network of different cancer cell lines, we observe that the proposed model (SANEpool) can detect the salient target gene patterns that cause the synergic drug combinations, which reveals the synergism mechanism in drug combination discovery.

\section{Other Related Work\label{sec2}}

\subsection{Graph Neural Networks}
Graph neural networks (GNNs) have revolutionized the field of learning with graph-structured data and empirically achieved the current state-of-the-art performance in various graph learning tasks, ranging from node classification and link prediction to graph classification. Broadly, GNNs \cite{schlichtkrull2018modeling,zhang2018end,ying2018hierarchical,verma2018graph,dai2016discriminative,niepert2016learning} follow a recursive neighborhood aggregation scheme where the node features from the neighborhood of each node are aggregated to update the node’s feature. Such frameworks allows GNNs to capture the graph topology as well as node features, hence unleashing the representation learning ability among graphs.

\subsection{an-cancer Biomarkers}
The genotype-oriented therapies for pan-caner biomarkers have been approved by the US Food and Drug Administration. These biomarkers amplify our knowledge of genomic profiling across various malignancies by revealing the prevalence of certain oncogenic alternations, hence playing important roles in drug combination discovery. According to a previous study \cite{yao2021pan}, 30$\%$ of recurrent alternations across tumor types from 10,000 patients with metastatic cancers are targetable, and various genotype-oriented therapies are detected based on genomic profiling. For instance, the neurotrophic receptor kinase (NTRK) family genes 1--3 were identified in various pediatric cancers. Then, a clinical trial of the Trk inhibitor larotrectinib demonstrated the antitumor activity and hence led to the usage of arotrectinib as treatment for cancers harboring NTRK~fusions.

\subsection{Machine Learning in Drug Synergy Prediction}
The drug synergy analysis is beneficial as it provides a useful resource for novel predicted drug combinations. However, manually discovering the synergism in practice is still challenging due to the high cost and the limited number of synergistic drug combinations approved by the Food and Drug Administration. Hence, the computational model shows huge potential to find the mechanism of synergy (MoS) in a biologically meaningful manner. Currently, various computational models, ranging from unsupervised learning models \cite{jin2011enhanced,wu2010systems,chen2016synergy} to supervised learning models~\cite{li2015large,xu2011exploring}, have been proposed for the purpose of predicting the synergy of drug combinations and have achieved expressive performance. Broadly, these computational models, such as DeepSynergy \cite{preuer2018deepsynergy} and Matchmaker \cite{kuru2021matchmaker}, take as input massive chemical descriptors of tested drug pairs and cell-line gene expression profiles and then use multi-layer-perceptron (MLP)-based deep learning models to predict the synergy score of drug combinations. Although these models effectively predict the synergy score of drug combinations, they are inherently not interpretable, while the interpretability is crucial for the real-world application. As the drug synergy has been reported to be largely determined by the biomolecular network topology \cite{yin2014synergistic}, many deep learning models, \linebreak  such as 
 DeepSignalingSynergy \cite{zhang2021synergistic} and IDSP \cite{dong2021interpretable}, incorporate the gene--gene interactions and gene--drug interactions into model design to allow the model make interpretable predictions that explain the underlying MoS.

\section{Methodology\label{sec3}}

\begin{figure}[H]
    \hspace{-1pt}\includegraphics[width=0.99\textwidth]{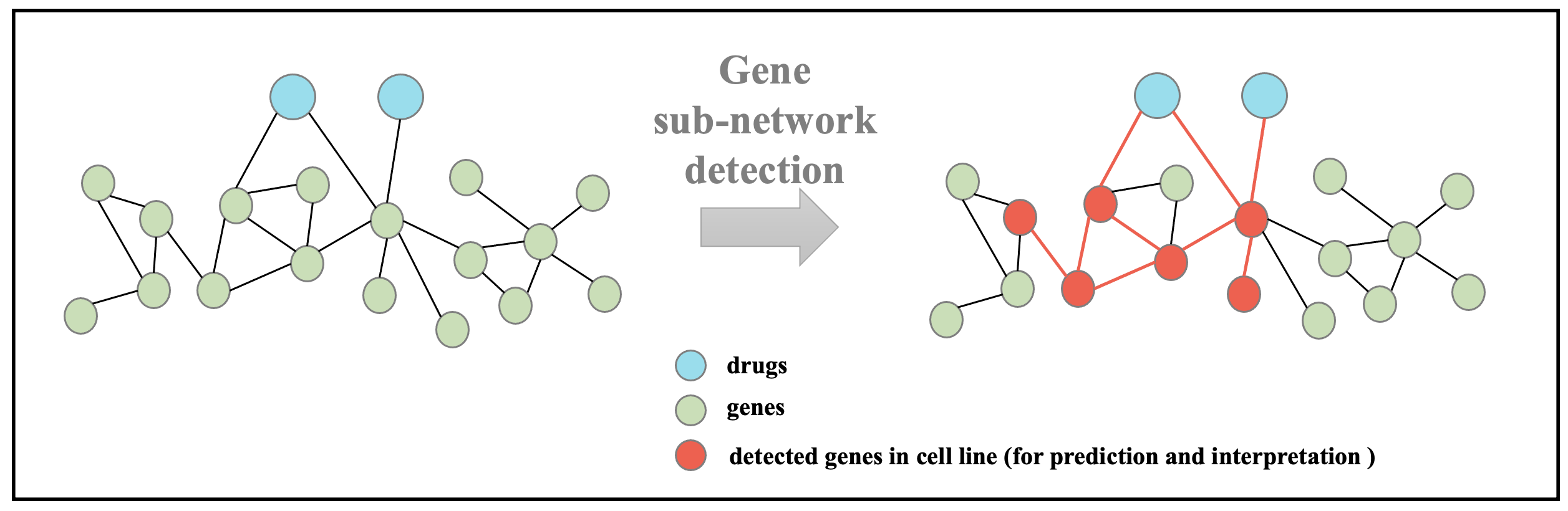}
    \caption{Overview of the problem formulation. The objective is to systematically detect gene sub-networks to predict the synergy score of the input drug pairs and to explain the drug combinatorial synergies across a large group of drug combinations and cell lines.\label{fig:tasks}}
    \end{figure}

\begin{figure}[H]
    \begin{adjustwidth}{-\extralength}{0cm}
\centering
    \includegraphics[width=0.99\linewidth]{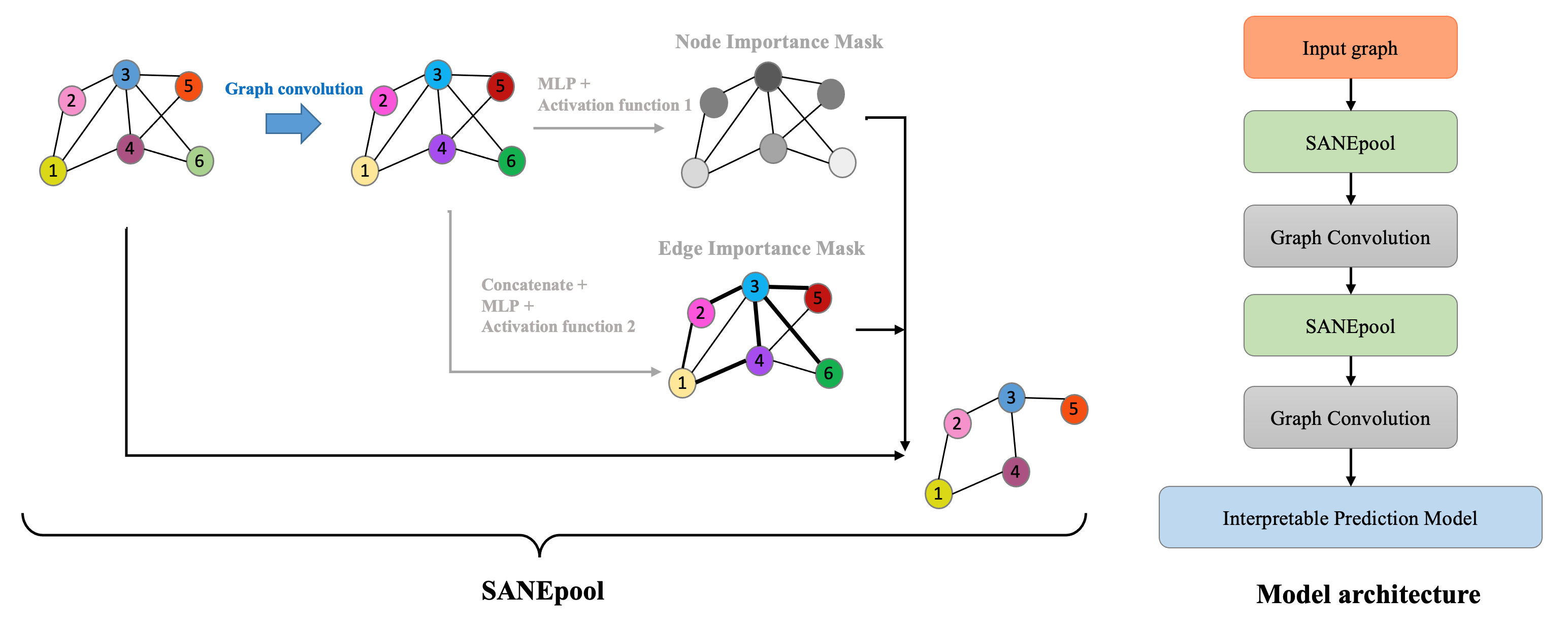}
\end{adjustwidth}
\vspace{-6pt}
    \caption{The 
 proposed SANEpool layer and the overall architecture. SANEpool layer incorporates node features and graph topologies through a GNN layer, and then the output is used to compute the attention score of nodes and edges, on the basis of which we can detect important nodes and edges through top-$\textit{k}$ 
 sorting or threshold filtering. The overall model takes a hierarchical graph pooling architecture. \label{fig:sanepool}}   
\end{figure}

\textls[-5]{In this section, we introduce the proposed graph pooling model, self-attention-based node and edge Pool (SANEpool), on the basis of which we develop an interpretable graph neural network to detect the biologically meaningful gene sub-network for synergic and interpretable drug combination prediction. The key point of SANEpool is to contain the attention score (importance) of nodes and edges though the node features and the graph topology; then, the attention scores make it possible to filter out less important (less relevant) nodes and edges for prediction. In Section \ref{sec3.1}, we introduce the problem formulation of the interpretable drug prediction task. In Section \ref{sec3.2}, we develop the mechanism of SANEpool, and the overall interpretable model architecture is described in Section \ref{sec3.3}.  The problem formulation and the architecture of SANEpool are illustrated in Figures~\ref{fig:tasks} and \ref{fig:sanepool},~respectively.}

\subsection{Problem Configuration\label{sec3.1}}
In this work, we study the molecular networks of cancer drug combination therapies in an inductive manner, where the tested drug pairs are unseen during the training phase. The molecular networks contain drugs and genes in the signaling pathways. The objective is to predict the synergic score of each drug pair based on the molecular network.  In order to make the prediction interpretable, SANEpool is proposed to detect the sub-gene network (i.e., the red gene nodes in Figure~\ref{fig:tasks} which consist of a subset of genes in signaling pathways that are most relevant to the synergistic effect of the drug pair. Then, the detected sub-gene network provides insight into the molecular mechanism of resistant or sensitive responses to cancer drug combinations. 

Let $G=(V,E)$ be the molecular network (graph), where $V$ is the node set that contains gene nodes and drug nodes, $E$ is the edge set that characterizes the interactions between nodes. For the notation convenience, we use $A$ to denote the adjacency matrix of the graph. Since the molecular graph is inherently undirected and has no self-loop, adjacency matrix A is a symmetric matrix.  We use $X \in R^{n \times h}$ to denote the input node feature, where $n$ is the number of nodes, and $h$ is the dimension of input features. Hence,~the graph can also be represented as the pair of the node feature and adjacency matrix such that $G=(X,A)$. Furthermore, we use $Z^t \in R^{n \times h_t} )$ to denote the node representation in layer $t$, where $h_t$ is the dimension of node representation.
\vspace{-9pt}
\vspace{-21pt}
\subsection{The Proposed SANEpool Model\label{sec3.2}}
The (self-)attention mechanism plays important role in various machine learning models, including natural language processing architectures \cite{parikh2016decomposable,devlin2018bert,yan2020does}, graph classification architectures \cite{Velickovic2018GraphAN}, sequential prediction algorithms \cite{cheng2016long}, adversarial learning models \cite{zhang2019self}, etc. The attention mechanism allows input features to be the criteria for the attention itself \cite{vaswani2017attention} and thus can distinguish the relative importance between features during the information aggregation process. In order to incorporate the node features and graph topologies in the attention scores for the nodes and edge pooling, we follow the idea of SAGpool \cite{lee2019self} and utilize a graph convolution layer to aggregate such information and to compute the attention feature matrix $H^t$.
\vspace{-6pt}
\begin{align}
    H^t = f(\tilde{D}^{-1}\tilde{A}Z^{t}\Theta^{t})
\end{align}
where $\tilde{A} = A + I$ is the adjacency matrix with added self-loops, $\tilde{D}$ is the corresponding diagonal degree matrix of $\tilde{A}$ such that $\tilde{D}_{ii} = \sum_{j=1}^n\tilde{A}_{ij}$, and $\textit{f}$ 
 is the activation function. The matrix $\Theta^{t} \in \mathbb{R}^{h_t \times h_{t+1}}$ is the trainable parameter to coordinate the attention score of nodes and edges, where $h_t$ is the feature dimension in layer $t$. Various graph convolution layers \cite{defferrard2016convolutional,niepert2016learning,kipf2016semi} have been proposed; these GNN formulas can be used as substitution of Equation (1). All GNN layers follows the same information aggregation framework; hence, the extracted attention features $H^t$  contain information of node features as well as graph~topologies.

Next, we discuss how to compute the attention scores of nodes and edges based on extracted attention features $H^t$. The node attention score is determined as the cosine of the angle between the attention feature vector and a trainable projection/parameter \mbox{vector $p^t$} of size $h_t$ (Equation (2)). \textbf{The attention score 
 $\textit{Att}$ measures the probability that gene nodes will cause the synergism of drug combinations.} Hence, we can sort node attention scores and adopt the top-$\textit{k}$ 
 selection technique to hierarchically select the gene sub-network for the purpose of synergism prediction, and such a process is formulated as
\vspace{-6pt}
\begin{align}
    \textit{Att}^{t}_{i} & = \frac{(p^t)^{T} H^{t}_{i}}{||p^t||^{\frac{1}{2}} ||H^{t}_{i}||^{\frac{1}{2}}} \\
    \textit{idx}^{t} &= \textit{top} (\textit{Att}^{t},k)
\end{align}

The core idea of the proposed SANEpool is to filter out ‘less important’ genes (and connections), and then only kept genes (and connections) are used to predict the synergism of drug combinations. Equation (2) indicates that the attention score $\textit{Att}$ is normalized and has a value in the set $[-1,1]$. Thus, a gene with larger attention score will be more likely to be kept in the top-$\textit{k}$ selection process (i.e., Equation (3)), indicating that the gene has a higher probability to be selected to predict/interpret the synergism of drug combination.

In the top node selection process (i.e., Equation (3)), when $k \in N$, we adopt the top-$\textit{k}$ node selection method as DGCNN \cite{zhang2018end}. On the other hand, we can also implement node selection method proposed by \cite{cangea2018towards} to retain a proportion of nodes when $k \in (0,1]$. Based on selected node index $\textit{idx}$, we can construct the graph downsampling as $\tilde{Z}^{t+1}= Z^t (idx,:)$ and $A^{t+1}= A^t (idx,idx)$. 

For general graph learning problems, it can be difficult to find the reasonable pre-defined indexes for nodes in graphs, as the problem is equivalent to the graph isomorphism problem, which is known to be NP-hard. However, for gene networks, since each gene at most appears once in each network, we can universally assign each possible gene (that appears in at least one network/graph in the data set) a unique (pre-defined) index. For instance, we can collect all possible genes and lexicographically sort their names, and then the order of genes in the sorting operation can be used as the index. Consequently, the pre-defined index is equivalent to the gene. It does not matter if we change the pre-defined index system once these indexes can injectively distinguish genes and are consistent among gene networks. The main advantage is to reduce the space complexity. The gene network usually contains lots of genes (i.e., a very large $n$), then, to store all possible edge pairs in the layer $t$, we require a tensor $\mathbb{T}$ with the shape of $\mathbb{R}^{n \times n \times 2 \times h_t}$,  \mbox{where $h_t$ is} the size of node representation vectors in layer $t$ and usually is selected from the set $\{32,64,128,256, 512\}$. Then an MLP is used to learn the edge attention matrix from $\mathbb{T}$.\linebreak In contrast, using pre-defined indexes can reduce the size of $\mathbb{T}$ to $\mathbb{R}^{n \times n}$, and no MLP is needed in this step. Hence, we can significantly shrink the model/algorithm complexity.

Let $G^{t+1}=(Z^{t+1},A^{t+1})$. Since the proportion of retained information (attention score) of nodes in $G^{t+1}$ are different, the connectivity strength between nodes can be different. Hence, we should also provide a consistent mechanism to characterize such bias. One intuitive approach is to apply the graph attention mechanism based on the extracted attention features  $H^t$:
\vspace{3pt}
\begin{equation}
    e^{t}_{i,j} = \textit{relu}(\textit{MLP}(H^{t}_{i} \ \lVert \ H^{t}_{j}))
\end{equation}
where the symbol $\lVert$ indicates the concatenation operation. Due to the universal approximation theorem \cite{hornik1989multilayer, hornik1991approximation}, such formulations can approximate any continuous function that measures the connectivity strength. However, a major limitation of this framework is the memory cost. For large-scale graphs, the memory usage to compute the attention score of edge may limit the practical ability of the proposed model. Luckily, the molecular network (graph) takes advantage that each gene node in the graph has a corresponding predefined index (i.e., the gene name), which serves as a canonical node order. Hence, we can directly model the interaction strength in each layer t through a trainable parameter matrix $W^{t}$. The the edge weight in the subgraph $G^{t+1}$ is trainable through the equation $A^{t+1}= A^{t+1} \circ W^t (idx^{t},idx^{t})$, where $\circ$ denotes the Hadamard product operation. The advantage of this formulation is discussed in Appendix D.

\subsection{The Overall Architecture\label{sec3.3}}

\subsubsection{Hierarchical graph pooling}
The overall architecture of the proposed interpretable model takes the hierarchical graph pooling structure \cite{cangea2018towards,lee2019self}. Figure~\ref{fig:sanepool} illustrates the overall architecture, and details are provided in Appendix E. The model stacks multiple SANEpool layers followed by a graph convolution layer to hierarchically extract a key sub-graph from the input graph. In other words, the proposed SANEpool layer is used to downsample the important sub-network (sub-graph). After the downsampling process, another GNN (graph convolution) layer is used to aggregate information based on sub-graph $G^{t+1}$ (Equation (5)) to update the node representation.
\vspace{5pt}
\begin{equation}
    Z^{t+1} = \textit{GNN} (Z^t[\textit{idx}^{t},:], A^{t+1})
\end{equation}
where $Z^t[\textit{idx}^{t},:]$ is the node representation matrix of the downsampled sub-graph in the $t$-th SANEpool layer. Then, the output of the last graph convolution layer is used for the prediction task. 

\subsubsection{Readout mechanism}
Inspired by IGMC \cite{zhang2019inductive}, the proposed model takes the node representation of two drug nodes to make prediction. The graph convolution framework indicates that such node representations encode the enclose rooted subtrees around the drug nodes in the pooled graph, hence representing the relations and interactions between drugs and genes. Ideally, we hope that the readout phase should be invariant to the drug node order, as the same drug pairs always have the same clinical performance regardless their order. Let  $u_1$, $u_2$ be the output of two drug nodes; then, the readout layer follows the factorization decoder in decagon \cite{kuru2021matchmaker}:
\begin{equation}
    \textit{score} = u^{T}_{1} D^{T} D u_2
    \label{equal:6}
\end{equation}
where $D \in \mathbb{R}^{t_{L},t_{L}}$ is the trainable parameter matrix in the decoder, where $t_{L}$ is the dimension of node representations in the last graph convolution layer $L$. The parameter matrix $D$ models the interaction effects between every two dimensions in drug representations \mbox{$u_{1}$ and $u_{2}$.}

It can be shown that the above readout function is invariant to the order of drugs. First, since the vectors/embeddings of two drugs,  $u_1$ and $u_2$, are generated by GNN message passing layers, which are invariant to the order of nodes/vertices in the input graphs, \mbox{$u_1$, $u_2$} will not change if we permute the order of drugs (even the order between two drugs and genes). Second, Equation (6) uses a symmetric function to compute the synergy score based on $u_1$ and $u_2$ and thus is permutation-invariant to the order of two drugs.

\subsection{Comparison to Related Works\label{sec3.4}}

\subsubsection{Comparison to other graph pooling models}
Both SANEpool and Sortpool \cite{zhang2018end} propose to sort nodes according to the structural role (i.e., `importance') of nodes in the graph. However, DGCNN is inherently flat, while SANEpool aggregate information in a hierarchical way. Thus, SANEpool is capable of capturing more informative global features for the downstream prediction task. On the other hand, SANEpool and Diffpool \cite{ying2018hierarchical} learn graph representation in a hierarchical way. However, Diffpool focuses on the relational analysis of node clusters, while SANEpool detects the critical sub-network based on the downstram tasks. Hence, SANEpool supports the pathway-based analysis in the drug combination prediction, thus providing interpretable results for healthcare.
\subsubsection{Comparison to other GNNs for drug synergy prediction}
The proposed SANEpool model, SDCNet \cite{zhang2022predicting}, and DeepDDS \cite{wang2022deepdds} share the same motivation that utilizes GNNs to capture the useful relational information of drugs and genes in the drug synergy prediction. However, they use GNNs to extract different types of relational information. The proposed SANEpool model encodes the important interactions between drugs and target genes in cell lines; SDCNet learns the cell-line-specific drug interactions, while DeepDDS encodes the molecular graph of each drug, which is generated by RDKit \cite{bento2020open} based on the drug's chemical structure. Consequently, compared to SANEpool, SDCNet and DeepDDS rely on more drug profile information. Furthermore, SDCNet is used for the simpler classification task which aims to predict the synergistic effects (0/1 classification) instead of accurately predicting the synergy score.




\section{Experiments\label{sec3}}
We evaluate the predictive ability of the proposed SANEpool by comparing the accuracy of the estimated synergy score against popular baselines. Furthermore, to illustrate the synergism detected by SANEpool, we also implement visual analytics and statistical analysis to show that the proposed SANEpool can detect significantly different causal gene sub-networks for synergic drug combinations and non-synergic drug combinations in each cell line.

\subsection{Dataset Description\label{sec4.1}}

We evaluate the effectiveness of the proposed SANEpool model on three drug-synergy-prediction datasets: NCI-DCD (NCI-Almanac-based Drug Combination Dataset), GDSC-SDD (Genomics-of-Drug-Sensitivity-in-Cancer \cite{jaaks2022effective} -based Single Drug Dataset) and O'Neil-DCD (O'Neil \cite{o2016unbiased}-based Drug Combination Dataset). 
Overall, these datasets take input molecular networks/graphs, which consist of drug combinations/pairs and genes in the target cell lines, where gene expressions are used as node features in the input molecular networks/graphs. The objective is to predict the score/synergy score of each drug/drug pair. In all datasets, the edges/interactions between drugs and genes are collected from the DrugBank database (version 5.1.5, 
released 3 January 2020
) \cite{wishart2018drugbank}, while the edges/interactions between genes are collected from the KEGG 
 (Kyoto Encyclopedia of Genes and Genomes) database \cite{kanehisa2000kegg} based on the physical signaling interactions from documented medical experiments. The  synergy score corresponding to each drug pair is computed as the average combo-score \cite{pan2017synthetic} with different doses on a given tumor cell line. The difference between datasets is the source of drug combinations and signaling pathways.

\subsubsection{NCI-DCD dataset}
The NCI-DCD ensemble genes from 46 well-known signaling pathways (45 “signaling pathways” + cell cycle) \cite{feng2021investigating} in KEGG. Drug combinations are collected from the DrugBank database \cite{wishart2018drugbank}, whose target genes are included in the aforementioned \mbox{46 signaling} pathways, and the combo-scores of drug combinations are available from the NCI Almanac dataset. We provide details of the $46 $ signaling pathways and $21$ selected FDA-approved drugs in Appendix \ref{sec:append2}. In summary, NCI-DCD contains $5658$ graphs/networks. Each graph/network has $1364$ genes and two drugs, while containing about 25,000 edges that connect genes and drugs.

\subsubsection{O'Neil-DCD dataset}
In O'Neil-DCD, signaling pathway information is also formulated based on the gene expression data of $1047$ cancer cell lines in the Broad Institute Cancer Cell Line Encyclopedia (CCLE) database \cite{barretina2012cancer}. Drug combinations and their corresponding synergy scores are collected from O'Neil datasets \cite{o2016unbiased}, whose target genes are included in KEGG database \cite{kanehisa2000kegg}. In summary, there are in total $4637$ graphs/networks, and each contains two drug nodes and $1823$ gene nodes.

\subsubsection{GDSC-SDD dataset}
In GDSC-SDD, signaling pathway information is formulated based on the gene expression data of $791$ cancer cell lines in the Broad Institute Cancer Cell Line Encyclopedia (CCLE) database, while the corresponding drug/cancer-cell-line response data set are available in the Genomics of Drug Sensitivity in Cancer (GDSC) database. In the experiment, there are in total 16,761 graphs/networks, and each contains a drug node and $969$ gene nodes. In the dataset, since there is only a single drug node in the input network/graph rather than a pair of drug nodes, we can not use Equation (\ref{equal:6}) as the readout function. Instead, we use a two-layer MLP that takes as input the learnt embedding of the drug node in this setting.

The three benchmark datasets (i.e., NCI-DCD, O'Neil-DCD, GDSC-SDD) contain different numbers of networks/graphs, where each drug or drug pair on each cell line represents one network/graph. In each network/graph, gene nodes use three features as the input/initial node features: gene expression value and two 0/1 indicators to indicate whether the gene is connected to two drugs, while drug nodes set all these values \mbox{as $-1$.} Then, when two networks correspond to different cell lines, the same gene will have different gene expression values as the initial node feature in these two networks. Furthermore, each network contains a gene if and only if (1) its gene expression data is available and\linebreak  (2) the gene is available in the KEGG dataset to track the interactions between genes. Then, since NCI-DCD, O'Neil-DCD, and GDSC-SDD have different resources of cell-line-based gene expression data (NCI-DCD dataset uses gene expression data from its website, and the gene expression data were collected from CCLE for the other two datasets\hl{)} 
 and drug (pair)/synergy data, networks in the different datasets will contain different sets of genes.

\subsection{Baseline Methods\label{sec4.2}}
 
In popular deep learning models for drug-synergy prediction, we select three popular baselines: DeepSynergy \cite{preuer2018deepsynergy}, DeepSignalingSynergy \cite{zhang2021synergistic}, and TransSynergy \cite{liu2021transynergy}. Compared to other methods, DeepSynergy uses additional drug profile features. To make a fair comparison, we mask out additional input features with $0$ embeddings.

We also compare SANEpool with six widely adapted GNN baselines. These baselines can be categorized into two types: (1) flat GNNs: Graph Attention Network (GAT)  \cite{velivckovic2017graph}, Deep Graph CNN (DGCNN) \cite{zhang2018end}, Graph Isomorphism Network (GIN)~\cite{xu2018powerful}, and graph convolutional network (GCN) \cite{kipf2016semi} and (2) popular graph pooling models: Diffpool \cite{ying2018hierarchical} and SAGpool \cite{lee2019self}. For GCN, GIN, and GAT, we stack three graph convolution layers with 64 output feature channels and concatenate sum-pooled features from the layers to generate the graph representation, which is then passed to an MLP to predict the graph~label. 

In the SANEpool model, we keep 200 nodes in the last layer and keep 90\% edges in each SANEpool layer. The SANEpool model takes two graph convolution layers and two SANEpool layers.
To provide robust model performance, we perform five-fold cross-validation and report the accuracy averaged over five folds and the standard deviation of validation accuracies across the five folds.

\subsection{Experimental Results\label{sec4.3}}

\subsubsection{Predictive performance}
In the experiment, we demonstrate the effectiveness of SANEpool in predicting the synergy score of drug/drug combinations. Table \ref{tab:table1} illustrates the experimental results. The experimental results indicate that SANEpool can accurately predict the synergy score of drug/drug combinations and achieve the state-of-the-art predictive performance among these competitive baselines. 

Furthermore, (1) we find that SANEpool significantly improves the performance over flat GNNs (i.e., GIN, GCN, GAT, DAGNN), which indicates that hierarchical graph representation learning technique in molecular networks can provide informative graph embedding with biologically meaning. (2) SANEpool outperforms other hierarchical graph pooling algorithms, and this observation indicates that incorporating edge information in the graph pooling is a potential future direction in the molecular network analysis, where graphs always have thousands of high-centrality nodes. 
\vspace{-3pt}
\begin{table}[H]
\tablesize{\footnotesize}
\caption{Performance 
 evaluation. Best results are highlighted.\label{tab:table1}}
\begin{adjustwidth}{-\extralength}{0cm}
\begin{tabularx}{\fulllength}{m{3cm}<{\centering}CCCCCC}
\toprule
\multirow{2}{*}{\vspace{-5pt}{\textbf{Model}}} & \multicolumn{2}{c}{\textbf{NCI-DCD}} & \multicolumn{2}{c}{\textbf{GDSC-SDD}} &  \multicolumn{2}{c}{\textbf{O'Neil-DCD}} \\
\cmidrule{2-7} 
 & \textbf{Pearson'r} \boldmath{$\uparrow$} & \textbf{MSE} \boldmath{$\downarrow$} &  \textbf{Pearson'r} \boldmath{$\uparrow$} & \textbf{RMSE} \boldmath{$\downarrow$} &  \textbf{Pearson'r} \boldmath{$\uparrow$} & \textbf{RMSE} \boldmath{$\downarrow$}\\
\midrule
DeepSynergy & 0.589 $\pm$ 0.022 & 47.742  $\pm$ 2.950 &  0.703 $\pm$ 0.014 & 0.0166  $\pm$ 0.0020 &   0.537 $\pm$ 0.021 & 187.56 $\pm$ 16.75\\
DeepSignalingSynergy & 0.631 $\pm$ 0.019 & 45.218 $\pm$ 1.889 &  0.744 $\pm$ 0.011 & 0.0143 $\pm$ 0.0012 &   0.598 $\pm$ 0.022 & 166.15 $\pm$ 19.56 \\
TransSynergy & 0.644 $\pm$ 0.023 & 46.219  $\pm$ 3.208 &  0.794 $\pm$ 0.022 & 0.0129  $\pm$ 0.0031 &   \textbf{0.615} $\pm$ \textbf{0.020} 
 & \textbf{160.19} $\pm$ \textbf{17.33}\\
\midrule
GIN & 0.565 $\pm$ 0.042 & 51.732  $\pm$ 5.636 &  0.716 $\pm$ 0.015 & 0.0155  $\pm$ 0.0017 &   0.550 $\pm$ 0.019 & 184.58 $\pm$ 17.06 \\
GCN &0.494 $\pm$ 0.049 & 58.585  $\pm$ 5.618 &  0.707 $\pm$ 0.014 & 0.0169  $\pm$ 0.0014 &   0.540 $\pm$ 0.024 & 187.84 $\pm$ 18.39\\
DAGNN & 0.509 $\pm$ 0.025 & 57.827  $\pm$ 3.174 &  0.638 $\pm$ 0.016 & 0.0198  $\pm$ 0.0018 &   0.431 $\pm$ 0.023 & 213.28 $\pm$ 20.19\\
GAT & 0.571 $\pm$ 0.031 & 50.995  $\pm$ 3.021 &  0.623  $\pm$ 0.013 & 0.0230  $\pm$ 0.0015 &   0.522$\pm$ 0.017 & 189.27 $\pm$ 18.51\\
SAGpool & 0.537 $\pm$ 0.031 & 53.125  $\pm$ 4.116 &  0.568 $\pm$ 0.011 & 0.0270  $\pm$ 0.0024 &   0.478 $\pm$ 0.016 & 197.69 $\pm$ 22.94\\
Diffpool & 0.577 $\pm$ 0.022 & 52.449  $\pm$ 3.155 &  0.658 $\pm$ 0.014 & 0.0186  $\pm$ 0.0039 &   0.517 $\pm$ 0.026 & 191.27 $\pm$ 18.31\\
\midrule
SANEpool (our model)  & \textbf{0.656} $\pm$ \textbf{0.016}& \textbf{44.352} $\pm$ \textbf{2.241} &  \textbf{0.825 } $\pm$ \textbf{0.009}& \textbf{0.0113} $\pm$ \textbf{0.0013} &   \textbf{0.614} $\pm$ \textbf{0.019}& \textbf{159.29} $\pm$ \textbf{23.01}\\
\bottomrule
\end{tabularx}
\end{adjustwidth}
\end{table}

\subsubsection{Interpretability}
The interpretability of deep learning models has been a major limiting factor for the use of these models in real-world drug-combination synergy analysis since most usage cases require explanations of the features used in the model. There is a natural trade-off between the interpretability and the accuracy of decision models in application, and hence it is critical to find the balance between them.  Currently, there are multiple deep learning models which take massive drug chemical structure information and predict the synergy score in a fully untransparent manner, such as DeepSynergy and MatchMarker. Although they can achieve expressive predictive performance, the lack of interpretability somehow limits the power of these models in real-world applications. Among the selected basslines, DeepSignalingSynergy is constructed based on the standard multiple layer perceptron model, and hence it is inherently not interpretable.  Similarly, GCN and GIN follow the basic neighborhood aggregation framework that aggregates information from the neighborhood of each node and then updates the node feature of the node based on the aggregated feature and node feature itself, while such aggregation processes are not interpretable either. In contrast, attention-based deep learning models, such as GAT and TransSynergy, can provide interpretable conclusions. For instance, GAT provides interpretability by measuring the connection strength between genes and graphs in the input biomolecular graphs through the attention mechanism, and we can analyze the effect of drug nodes based the connection strength. Analogous to these interpretable models, in the next section, we will show that the interpretability of the proposed SANEpool model comes from its ability of computing the `synergic importance'  of each gene. The `synergic importance' is computed as the expectation of its effect on the synergic score of drug combinations. Specifically, the effect can be measured by whether it is used in the synergic score prediction process (0/1 variable, i.e., whether the gene is selected by SANEpool model) or we can multiply the 0/1 variable with the synergy score.  In this paper, we use the former definition. Hence, the `synergic importance'  can be used to detect genes with closer correlations to synergic drug combinations in each cell line by selecting genes whose `synergic importance' is larger than a given threshold. 

\subsection{Statistical Analysis and Visualizations}

\begin{figure}[H]
    \includegraphics[width=0.96\textwidth]{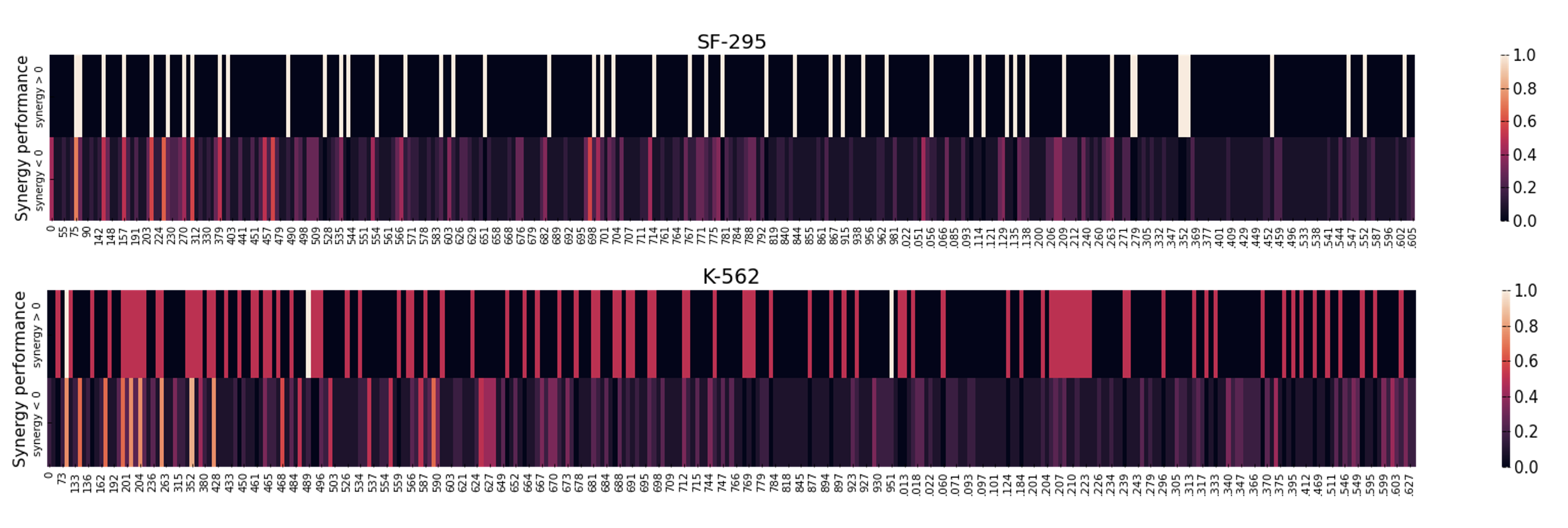}
\vspace{-3pt}
    \caption{Synergic/non-synergic importance scores of all genes in cell line SF-295 and cell line K-562. X-axis is the gene index. The same genes in different cell lines share the same index.  \label{fig:fig4}}  
\end{figure}

In the experiment, we implement statistical analysis and visual analytics to reveal the interpretability of the SANEpool model. Here we use NCI-DCD as an example. 
Previous works \cite{jaaks2022effective, hosseini2023ccsynergy} have emphasized that drug synergy is highly cell-line-specific/context-specific. Hence, we perform the cell-line-specific analysis. For each cancer cell line, there are multiple drug combinations targeting the cell line, some of which are synergic, while others are not synergic. For each drug combination, SANEpool can select top $\textit{k}$ genes for the prediction. Hence, we define the synergic/non-synergic importance score of a gene as the proportion of the gene is used by the SANEpool model in the prediction when input drug pairs are synergistic/non-synergistic. For instance, if a gene node is never detected by any synergistic drug combination that targets on a specific cancer cell line, then its synergic importance score is 0 and is never used to predict the synergy score of synergic combinations. Figure~\ref{fig:fig4} compares these scores computed by SANEpool, and it illustrates that 
the patterns of synergic importance scores and non-synergic importance scores across genes are different in each cell line.

Next, we should decide whether the detected gene sub-networks for synergic drug combinations are significantly different from those of non-synergistic drug combinations. In order to do so, we can compare the distribution of the synergic importance scores on genes and the distribution of the non-synergic importance scores. If these two distributions are significantly different, we can infer that the detected gene sub-networks are also significantly different. Hence, we implement the Kolmogorov--Smirnov test (K-S test) to compare these distributions. In the K-S test, the null hypothesis assumes that two distributions are the same, and it computes a $D$ statistic as well as a \emph{p}-value corresponding to the $D$ statistic. Then, we reject the null hypothesis if the \emph{p}-value is less than the significance level (0.05). We provide details of the K-S test and relevant statistics in Appendix \ref{sec:append3}. We implement the cell-line-based test. The cell-line based K-S test results are provided in Table \ref{tab:table2}, and it shows that the K-S test is significant for each cell line. Consequently, we use the difference in the synergic importance score and non-synergic importance score of each gene to determine whether the gene is selected in the core gene sub-network. That is, the gene is obtained in the core gene sub-network if the computed value is above a given threshold, like $0.1$. 
\vspace{-9pt}
\vspace{-11pt}
\begin{table}[H]
\tablesize{\scriptsize}
\caption{Cell 
-line-based K-S test results.\label{tab:table2}}
\begin{adjustwidth}{-\extralength}{0cm}
\begin{tabularx}{\fulllength}{m{1.7cm}<{\centering}CCCCCm{2cm}<{\centering}Cm{2cm}<{\centering}C}
\toprule

\textbf{Cell Line} & \textbf{\emph{p} Value} & \textbf{Cell Line} & \textbf{\emph{p} Value}  & \textbf{Cell Line} & \textbf{\emph{p} Value}  & \textbf{Cell Line} & \textbf{\emph{p} Value}  & \textbf{Cell Line} & \textbf{\emph{p} Value} \\
 
\midrule
 UACC-62 & 
 $<0.001$
 & NCI-H522 &  $<0.001$  & HT29 & 0.002  & MDA-MB-435 &  $<0.001$  & A549/ATCC & 0.003\\
 
 OVCAR-8 &  $<0.001$  & HOP-62 & 0.003  & HCT-15 &  $<0.001$  & RPMI-8226 &  $<0.001$  & MDA-MB-231/ATCC &  $<0.001$ \\
 
 OVCAR-3 &  $<0.001$  & HS 578T & 0.003  & UO-31 &  $<0.001$  & BT-549 & 0.005  & UACC-257 &  $<0.001$ \\
 
 LOX IMVI & 0.003  & SW-620 &  $<0.001$  & MCF7 &  $<0.001$  & NCI-H460 &  $<0.001$  & EKVX &  $<0.001$ \\
 
 HOP-92 &  $<0.001$  & SF-268 &  $<0.001$  & K-562 & 0.007  & T-47D & 0.002  & MDA-MB-468 & $<0.001$ \\
 
 MALME-3M &  $<0.001$ & SK-MEL-5 &  $<0.001$  & SF-295 & 0.004  & NCI-H23 &  $<0.001$  & OVCAR-4 & 0.002 \\
 
 SF-539 &  $<0.001$  & U251 &  $<0.001$  & PC-3 & 0.005  & CAKI-1 & 0.007  & HCT-116 &  $<0.001$ \\

IGROV1 &  $<0.001$  & SK-OV-3 & 0.006  & A498 &  $<0.001$ & NCI-H322M &  $<0.001$  & ACHN &  $<0.001$ \\

HL-60(TB) & 0.005  & KM12 & <0.001  & NCI-H226 & <0.001  & SK-MEL-28 & <0.001  & DU-145 & 0.004 \\
\bottomrule
\end{tabularx}
\end{adjustwidth}

\end{table}

In addition to the statistical analysis, we also use heat maps to show the difference between detected sub-networks of synergic drug combinations and non-synergic drug combinations. Here, we provide examples on cell line SF-295 and cell line K-562 in Figure~\ref{fig:fig4}, and more examples are provided in Appendix \ref{sec:append}. In the heatmap, the values (synergy $>$ 0, \mbox{synergy $<$ 0}) assigned to each gene are the cell-line-based synergic importance and non-synergic importance scores, which indicate the proportion that the gene is included in the gene sub-network (detected by SANEpool) of synergic drug combinations (non-synergic drug combinations) targeting the cell line. The synergy value of a gene measures the possibility that the information of the gene, such as the gene expression and gene copy number, contributes to the prediction process of the drug synergy score. 
Hence, the difference between cell-line-based synergy value (estimated probability that the gene is involved in the detected sub-network of drug combinations with synergy $>$ 0) and cell-line-based non-synergy value (estimated probability that the gene is involved in the detected sub-network of drug combinations with synergy $<$ 0) can reflect the synergic performance of the gene. That is, a larger difference indicates the gene contributes more to the synergic drug combination than non-synergic drug combinations. For instance, for cell line SK-562, the top 10 detected genes are SIN3A, ETS2, WNT10B, SLC8A1, MTOR, KLF2, RGS2, SESN3, NRG1, TNFRSF11A. Furthermore, these heatmaps also show that the difference between the detected genes of synergic drug combinations and of non-synergic drug combinations are significant.

\begin{figure}[H]
 s   \includegraphics[width=12cm]{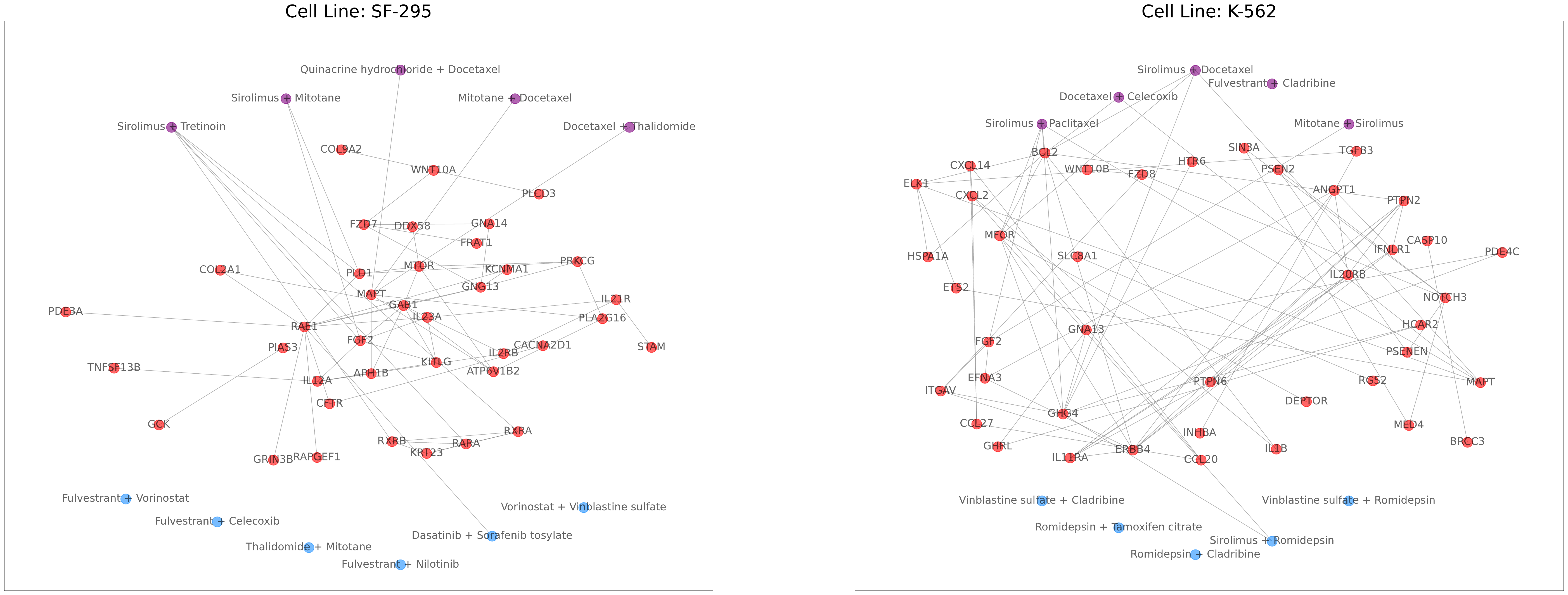}
    \caption{Visualization 
 of drug--gene interactions on selected cell lines. In these graphs, red nodes represent genes in the subnetwork of all synergic drug combinations on the cell line,  {purple nodes} 
 are synergic drug pairs, blue nodes are randomly non-synergic drug pairs.    \label{fig:fig5}}
\end{figure}

\begin{figure}[H]
    \includegraphics[width=12cm]{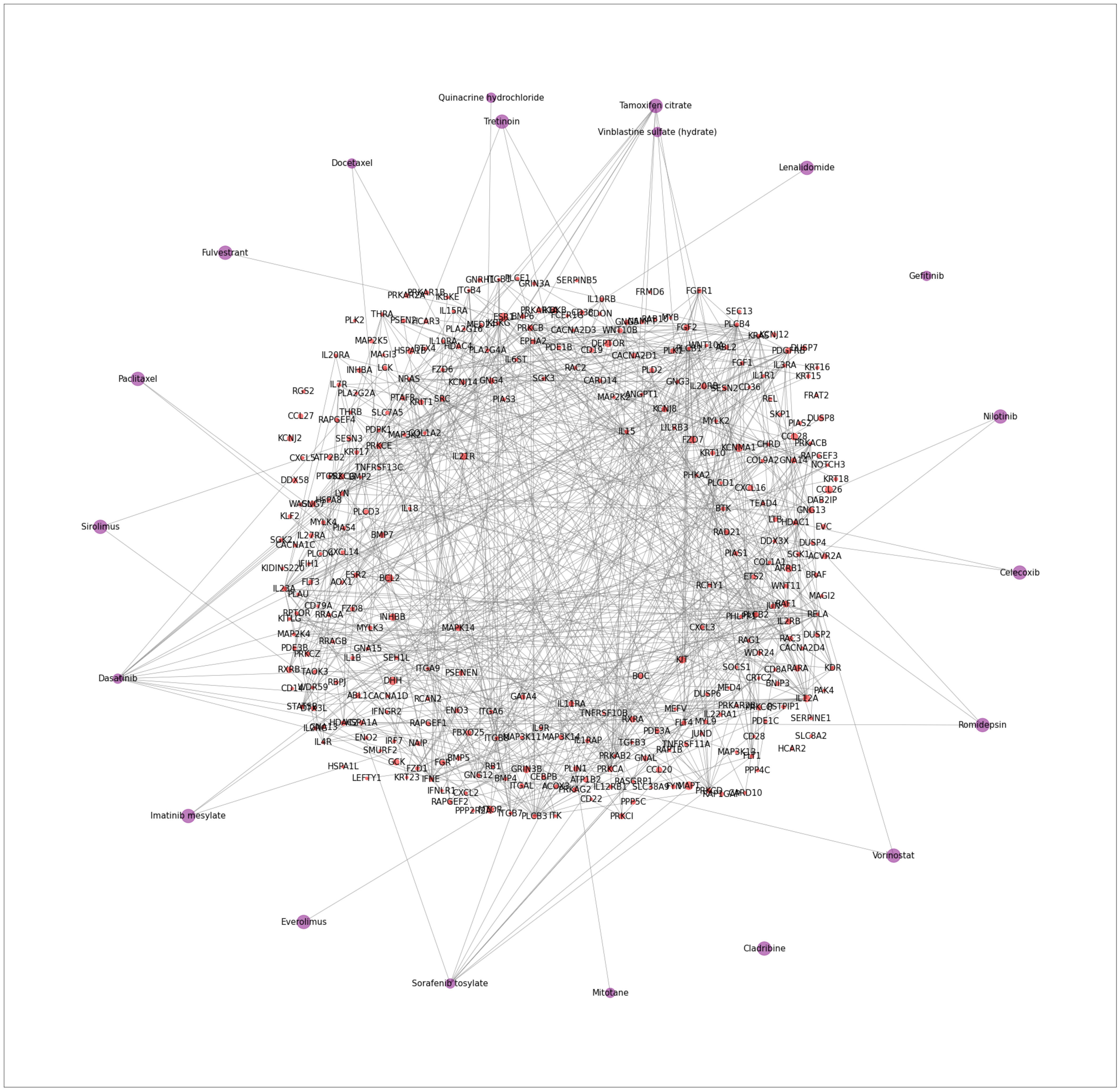}
    \caption{{The 
} figure describes the interactions of synergic drugs combinations and genes in the detected core gene sub-network. Synergic drug nodes are visualized as purple nodes, while red nodes represent genes in the detected core gene sub-network.\label{fig:fig6}}    
\end{figure}

Furthermore, we also visualize the interactions of drugs and genes in the detected core gene sub-network in Figures~\ref{fig:fig5} and \ref{fig:fig6} to show that synergic drug combinations are more likely to target on the detected core gene sub-networks in each cell line. Figure~\ref{fig:fig5} focuses on specific cell lines, while Figure~\ref{fig:fig6} combines all cell lines and drug combinations.
For each cell line, we plot all genes (i.e., red nodes) in the sub-networks (detected by SANEpool) of synergic drug combinations and then randomly sample synergic drug combinations (i.e., purple nodes) and some non-synergic drug combinations (i.e., blue nodes). Figure~\ref{fig:fig5} illustrates that drugs in the non-synergic combinations are very unlikely to target on the core gene sub-network in each example cell line (e.g., SF-295, K-562).  Figure~\ref{fig:fig6} indicates that this observation can be extended to other cell lines.
\vspace{-6pt}


\section{Conclusions\label{sec4}}
 In this paper, we have proposed an interpretable GNN architecture called SANEpool (self-attention-based node and edge pool) to predict the synergy score of drug combinations and to investigate the underlying mechanism of the synergy (MoS) by detecting salient molecular sub-networks. For each cell line and each drug combination, SANEpool can detect a specific sub-network, and SANEpool evaluates the contribution of each gene to synergic drug combinations based on all detected (cell-line-based) sub-networks. Hence, cell-line-specific essential signaling gene targets are identified by SANEpool. Furthermore, our observation also indicates that most synergistic drug combinations inhibit the core signaling network detected by SANEpool. The current work is limited by the number of drug combinations and cell lines. In future work, more drug combination datasets with multi-omic data will be integrated to uncover the mechanism of the synergy of effective drug combinations.    


\vspace{6pt} 



\authorcontributions{All authors have read and agreed to the published version of the manuscript. Conceptualization, Zehao Dong, Yixin Chen, and Fuhai Li; Methodology, Zehao Dong and Fuhai Li; software, Zehao Dong; validation, Zehao Dong; data curation, Heming Zhang, Philip Payne and Fuhai Li; writing---original draft preparation, Zehao Dong; writing---review and editing $\&$ editing, Zehao Dong, Heming Zhang, Yixin Chen and Fuhai Li; visualization, Zehao Dong; supervision, Yixin Chen, Philip R.O. Payne and Fuhai Li; project administration, Yixin Chen, Philip R.O. Payne and Fuhai Li; funding acquisition, Yixin Chen, Philip R.O. Payne and Fuhai Li. 
}

\funding{This work is partially supported by the Children’s Discovery Institute (CDI) M-II-2019-802, and NLM 1R01LM013902-01A1 to Fuhai 
 Li.}

\institutionalreview{Not applicable. 
}

\informedconsent{Not applicable.  
}

\dataavailability{Our source codes, which include datasets and implementation pipelines for reproducibility, are available at \url{https://github.com/zehao-dong},  (accessed 31 August 2023). 
} 

\conflictsofinterest{The authors declare no conflict of interest. 
} 


\appendixtitles{yes} 
\appendixstart
\appendix
\section{Cell-Line Based Visualization Results\label{sec:append}}
\begin{figure}[H]
     \includegraphics[width=13cm]{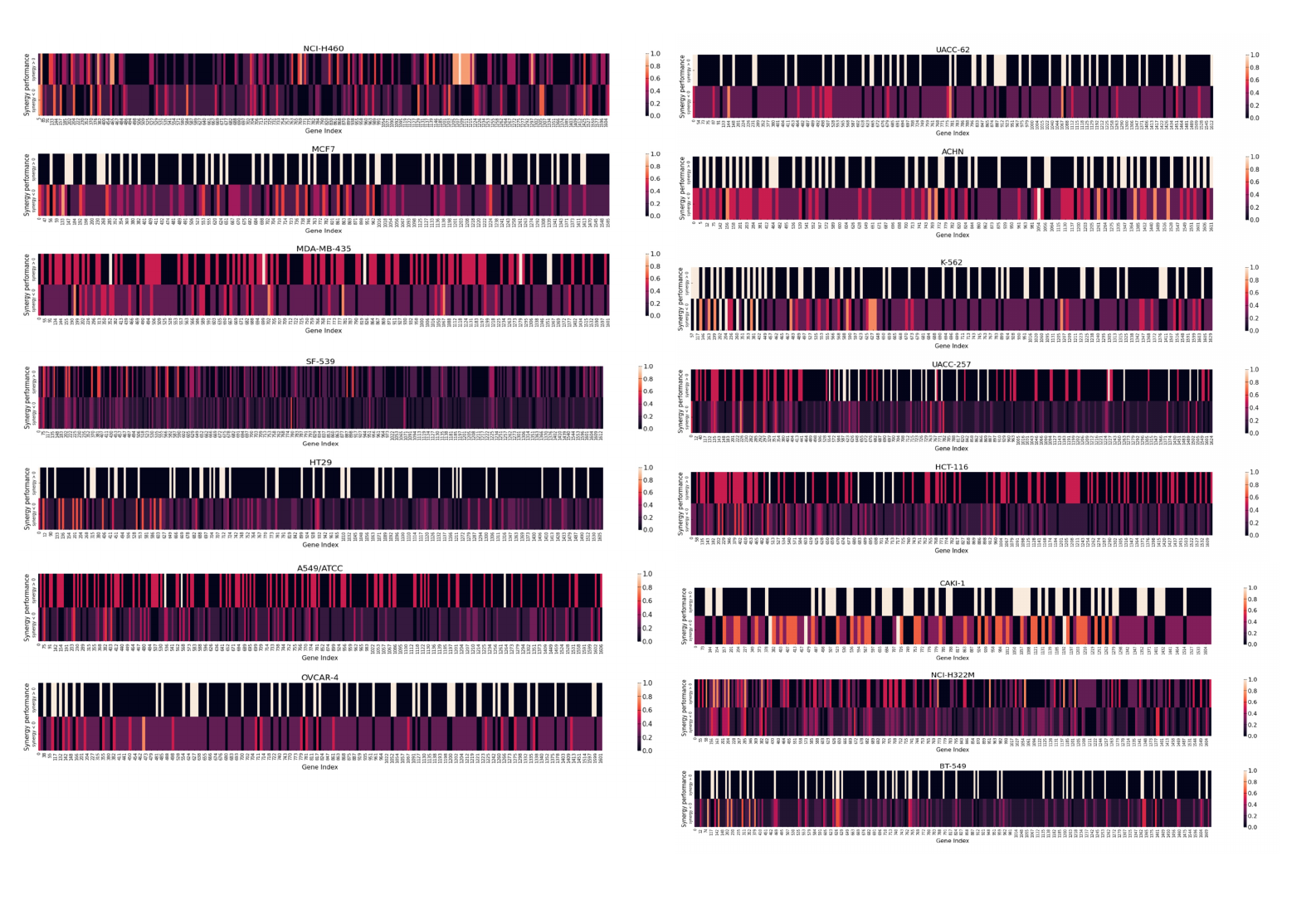}
\vspace{-21pt}
    \caption{Comparison 
 of gene synergy values in all cell-lines.\label{fig:fig10}}
    \end{figure}

In this section, we provide the assigned synergy value (estimated probability that the gene is involved in the detected sub-network of drug combinations with synergy > 0) and assigned non-synergy value (estimated probability that the gene is involved in the detected sub-network of drug combinations with synergy > 0) of all genes for each cell line. As Figure~\ref{fig:fig10} illustrates, the synergy value of a gene, which measures its importance in the prediction of synergy score, shows a significant difference in each cell line between synergic drug combinations (synergy > 0) and non-synergic drug combinations (synergy < 0).
\vspace{-13pt}

\section{Cell Lines and FDA Approved Drugs\label{sec:append2}}

This section introduces the cell lines and drugs used in the dataset NCI-DCD. For dataset O'Neil-DCD and GDSC-SDD, the relevant information is available in previous works \cite{o2016unbiased, jaaks2022effective}.

Signaling pathways used to formulate the input graphs: MAPK, FoxO, TGF-beta, \mbox{T-cell} receptor, Adipocytokine, ErbB, Sphingolipid, VEGF, B-cell receptor, Oxytocin, Ras, Phospholipase D, Apelin, Fc epsilon RI, Glucagon, Rap1, p53, Hippo, TNF, Relaxin, Calcium, mTOR, Toll-like receptor, Neurotrophin, AGE-RAGE, cGMP-PKG, PI3K-Akt, NOD-like receptor, Insulin, Cell cycle, cAMP, AMPK, RIG-I-like receptor, GnRH, Chemokine, Wnt, \mbox{C-type} lectin receptor, Estrogen, NF-kappa B, Notch, JAK-STAT, Prolactin, HIF-1, Hedgehog, IL-17, thyroid hormone. 

FDA approved drugs used to formulate the input graphs: Celecoxib, Gefitinib,\linebreak Quinacrine  hydrochloride, Tretinoin, Cladribine, Imatinib mesylate, Romidepsin,\linebreak  Vinblastine sulfate (hydrate), Dasatinib, Lenalidomide, Sirolimus, Vorinostat, Docetaxel,\linebreak  Mitotane, Sorafenib tosylate, Thalidomide, Everolimus, Nilotinib, Tamoxifen Citrate, Paclitaxel, Fulvestrant. 

\section{Details of K-S Test\label{sec:append3}}

Here, we introduce more details of the cell-line-based K-S test. The objective is to test whether the detected sub-gene network (by SANEpool model) is significantly different for synergic drug combinations (synergy score $>$ 0) and non-synergic drug combinations (synergy score $<$ 0). It can be difficult to directly test the hypothesis. However, since SANEpool sorts the attention score, Att, of genes to select gene in the `core sub-gene network', each gene has different probability to be selected by synergic and non-synergic drug combinations. Hence, we propose to compare the distribution of selected genes for synergic and non-synergic drug combinations in each cell line. Next, we need some observations to perform the K-S test. For each cell line, suppose that there are N 
 synergic and M non-synergic drug combinations on the cell line. Then, for synergic drug combinations, each observation will randomly sample $0.8 \times N $ synergic samples, and then use the proportion of each gene included in the detected sub-gene network (by SANEpool) as its probability of being selected. Then we get an observation $\textit{ps}_{i}$, which is a vector of probability of being selected by synergic drug combinations for all genes. Similarly, we can obtain an observation $\textit{pns}_{i}$ from $0.8 \times M$ non-synergic samples. Then, we sample 10 observations for both synergic and non-synergic drug combinations: $\textit{ps}_{i}$, $\textit{pns}_{i}$ for $i=1,2, ..., 10$. After that, we simply need to combine these observations and compute the empirical cumulative distribution function (cdf) F. Then, we can perform the two-sample K-S test (i.e., $D_n = \max(|F_{exp} - F_{obs}|)$),\linebreak  where $F_{exp}$ and $F_{obs}$ are the empirical cumulative distribution functions computed based on $\textit{ps}_{i}$ and $\textit{pns}_{i}$.

\section{Advantage of Using Pre-Defined Order of Genes in SANEpool
\label{App4}}

 For general graph learning problems, it can be difficult to find reasonable pre-defined indexes for nodes in graphs, as the problem is equivalent to the graph isomorphism problem, which is known to be NP-hard. However, for gene networks, since each gene at most appears once in each network, we can universally assign each possible gene (that appears in at least one network/graph in the data set) a unique (pre-defined) index. For instance, we can collect all possible genes and lexicographically sort their names, and then the order of genes in the sorting operation can be used as the index. Consequently, the pre-defined index is equivalent to the gene. It does not matter if we change the pre-defined index system once these indexes can injectively distinguish genes and are consistent among gene networks. The main advantage is to reduce the space complexity. Since the gene networks usually contain lots of genes (i.e., a very large n), then, to store all possible edge pairs in the layer t, we require a tensor T with the shape of $R^{n \times n \times 2 \times h_t}$,  where $h_t$ is the size of node representation vectors in layer t and usually is selected from 
 $[32,64,128,256]$. And then, an MLP is used to learn the edge attention matrix from T.  In contrast, using pre-defined indexes can reduce the size of T to $R^{n \times n}$, and no MLP is needed in this step. Hence, we can significantly shrink the model/algorithm complexity.

\section{Details of the Overall Model Architecture 
\label{App5}}
The overall model architecture is composed of SANEpool layer $\to$ graph convolution layer $\to$ batch normalization layer $\to$ SANEpool layer $\to$ graph convolution layer $\to$ batch normalization layer$\to$ readout layer (Equation (6)). Following this flow, the proposed model will output a predicted synergy score value for each input network/graph. In addition to the SANEpool layers and graph convolution layers, batch normalization layers are used to regularize the model training and to avoid overfitting.

\begin{adjustwidth}{-\extralength}{0cm}

\reftitle{References}


\PublishersNote{}
\end{adjustwidth}
\end{document}